\documentclass{article}

\usepackage{arxiv}

\usepackage[utf8]{inputenc} 
\usepackage[T1]{fontenc}    
\usepackage{hyperref}       
\usepackage{url}            
\usepackage{booktabs}       
\usepackage{amsfonts}       
\usepackage{nicefrac}       
\usepackage{microtype}      
\usepackage{lipsum}		
\usepackage{graphicx}
\usepackage[square,numbers]{natbib}
\usepackage{doi}

\title{Sequence-based deep learning antibody design for in silico 
antibody affinity maturation}

\date{Februrary 17, 2021}	

\author{ {\hspace{1mm}Yue Kang}, {\hspace{2mm}Dawei Leng},\href{https://orcid.org/0000-0002-9157-2199}  {\hspace{2mm}\includegraphics[scale=0.06]{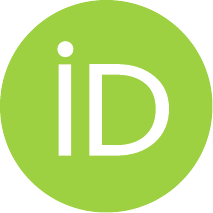}Jinjiang Guo*}, \href{https://orcid.org/0000-0001-9321-8348} {\hspace{2mm}\includegraphics[scale=0.06]{orcid.pdf} Lurong Pan*}\\
	\emph{AIDD Group}\\
	Global Health Drug Discovery Institute, Beijing, China \\
	\texttt{yue.kang@ghddi.org},  \texttt{jinjiang.guo@ghddi.org},  \texttt{lurong.pan@ghddi.org}  \\
}

\renewcommand{\headeright}{}
\renewcommand{\undertitle}{}
\renewcommand{\shorttitle}{Sequence-based deep learning antibody design for in silico 
antibody affinity maturation}

\hypersetup{
pdftitle={A template for the arxiv style},
pdfsubject={q-bio.NC, q-bio.QM},
pdfauthor={David S.~Hippocampus, Elias D.~Striatum},
pdfkeywords={First keyword, Second keyword, More},
}

\begin{document}
\maketitle

\begin{abstract}
	Antibody therapeutics has been extensively studied in drug discovery and development within the past decades. One increasingly popular focus in the antibody discovery pipeline is the optimization step for therapeutic leads. Both traditional methods and in silico approaches aim to generate candidates with high binding affinity against specific target antigens.  Traditional in vitro approaches use hybridoma or phage display for candidate selection, and surface plasmon resonance (SPR) for evaluation, while in silico computational approaches aim to reduce the high cost and improve efficiency by incorporating mathematical algorithms and computational processing power in the design process. In the present study, we investigated different graph-based designs for depicting antibody-antigen interactions in terms of antibody affinity prediction using deep learning techniques. While other in silico computations require experimentally determined crystal structures, our study took interest in the capability of sequence-based models for in silico antibody maturation. Our preliminary studies achieved satisfying prediction accuracy on binding affinities comparing to conventional approaches and other deep learning approaches.  To further study the antibody-antigen binding specificity, and to simulate the optimization process in real-world scenario, we introduced pairwise prediction strategy. We performed analysis based on both baseline and pairwise prediction results. The resulting prediction and efficiency prove the feasibility and computational efficiency of sequence-based method to be adapted as a scalable industry practice.
\end{abstract}


\section{Introduction}
\label{sec:headings}
The interaction between antibody and antigen is achieved through the non-covalent bonds between amino acid residues on the interacting surfaces, i.e. the interacting contacts (ICs). A high binding affinity reflects a stable antibody-antigen complex \cite{ref12}, which requires particular amino acids composition as well as conformation upon binding. Researchers have previously reported that at 80 $\%$ of the binding free energy is clustered on a very limited number of interacting residues\cite{ref14}. Specifically, the complementarity-determining region on the antibody, or CDR, is considered to be what allow antibody to bind to a target with high specificity \cite{ref10,ref11, ref12}. Although residues on non-interacting surface do not bind with antigen directly, studies have demonstrated that non-interacting surface (NIS) also plays an important role in binding affinity by long-range electrostatic contributions and surface–solvent interactions \cite{ref9,ref13} 

Traditional in-silico approaches compute the antibody-antigen binding strength based on atom chemical properties such as polarity and charge\cite{ref9}. With the fast development in the field of machine learning and deep learning research, algorithms and techniques are explored towards the application of antibody development. \cite{ref5} studied the prediction of interacting contacts and performed binding prediction using surface-based geometric features; \cite{ref4} employed topology-based network tree for binding affinity change prediction based on complexes’ 3D structures. 	Efforts to improve antibody-antigen modeling also include deep learning analysis using sequence data only.  \cite{ref3} trained long-short-term memory models on in silico sequence library for antigen-specific affinity predictions. In \cite{ref8}, mutations within CDR-H3 region are modeled and optimized with respect to different antigens by employing ensembled CNN models.In the present study, we proposed GNN-based modeling strategies for antibody maturation by utilizing sequence data only. Base on the Hag-Net graph neural network structure\cite{ref1}, we investigated the contributions of interacting contacts (ICs), non-interacting surface (NIS), as well as antibody-based features in terms of affinity prediction through different modeling designs. The proposed approaches achieved satisfactory classification and regression performance in both five-fold cross-validations and out-of-distribution analysis, and outperformed conventional in-silico analysis on AB-Bind dataset.

\section{Data and Methodology}

\subsection{Data Collection}
Antibody-Bind (AB-Bind) is a manually curated and organized database that includes 1101 mutants across 32 complexes, with experimentally determined binding affinities upon mutations \cite{ref22}. The binding affinities for mutated variants are represented by the change in free energy ($\Delta\Delta G$) of binding in kal/mol.  The study also present binging affinity predictions using the bASA, Rosetta, dDFIRE, DFIRE, STATIUM, FoldX, and Discovery Studio scoring functions\cite{ref22}. The presented methods were further evaluated in terms of their binary classification performance, where mutated variants are categorized as improved vs weakened binders. In the presented study, we follow the same protocols and construct both regression and classification tasks based on AB-Bind dataset. For more details on AB-Bind data and reported computations, the reader is referred to \cite{ref22}.

\subsection{Methodology}
\subsubsection{Hag-Net network structure}
The graph neural networks \cite{ref2} are neural network architectures that designed specifically to cope with graph data. Nodes in the graph are designed to learn an embedding containing information about their associated neighborhoods. The embeddings can be utilized as characteristic features for node labeling, edge prediction as well as graph representation with proper readout/pooling methods. Traditional machine learning applications use preprocessing techniques to map graph structure data into simpler representations, whereas the graph neural networks extends existing neural network methods in order to retains the important information such as the topological dependency of information on each node	. The intrinsic design of GNN makes it well-suited to study the molecular and biological interactions and properties, we therefore leverage the power of graph representations to improve the antibody-antigen interaction modeling in the present study.

Specifically, we employed the Hag-Net graph neural network for antibody affinity change prediction. The Hag-Net network structure is invented and pioneered by the AIDD Group \cite{ref1} of Global Health Drug Discovery Institute. As a variant of spatial based graph neural networks, Hag-Net aggregates features by taking both summation heterogeneous graph based deep learning for biomedical network link prediction and maxima of neighbouring features \cite{ref1}, and therefore increase effect information propagation from shallow layers to deep layers. For graph G = (V, E) with $v, u \in V$ and $e_{uv} \in E$, the generic layer formulation with heterogeneous aggregations is as:

\begin{equation}
h_v  =  \psi (C(h_v,	\bigoplus_{i=1}^{M-1} \phi_i(A_{i,u\in N(v)}( \{ (h_v,h_u,e_{uv})\}))))
\end{equation}

Eq(1) describes the update method for center node $h_v$. $A$ represents M different aggregation operators, while  $\bigoplus$ represents the merge operator for M neighborhood aggregation results. $C(·) $ updates $h_v$ with the merged aggregation result. $ \psi$ and $\phi_i$ are linear/non-linear transform functions.

In this application, we utilized the graph-representation of  Hag-Net for label prediction,which requires the READOUT function in order to aggregate node features to form the graph-level output:

\begin{equation}
h_G  =  \psi(	\bigoplus_{i=1}^{M-1} \phi_i(A_{i,v\in V}(\{(h_v\})))
\end{equation}

The final classifier layer (consists of several dense layers) takes the graph-level output as input and outputs model prediction for assigned tasks.

\subsubsection{Baseline study:\textit{ binding affinity change ($\Delta\Delta G$}) prediction}
\begin{figure}
	\centering
	\includegraphics[width = 16 cm]{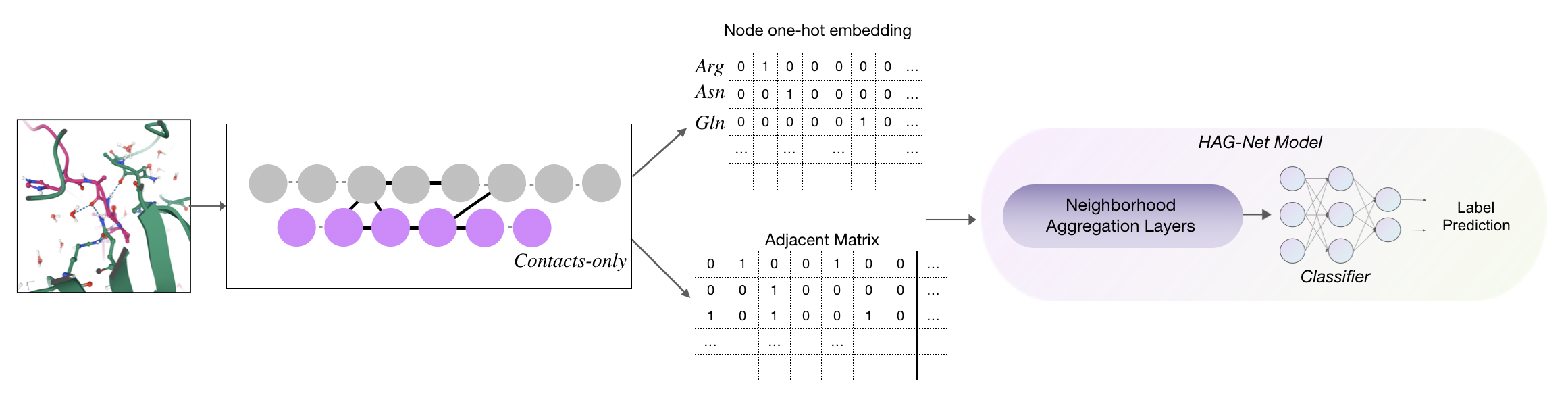}
	\caption{Contacts-only modeling for weakened vs improved binder classification based on Hag-Net network structure. }
	\label{fig:fig1}
\end{figure}

We performed antibody-antigen complex affinity prediction based on sequence data with Hag-Net network structure. In the proposed modeling approach, each node represents a single residue, and its edges represent connections with associated/interacting residues. This modeling approach represents each antibody-antigen complex as a single graph structure that corresponds to the amino acid sequences of antibody and antigen. The resulting graph was then converted to an input matrix by one-hot encoding, where each row represents a specific residue, as well as an adjacent matrix, where each $1$ represents connection between amino acids in two positions. Thus, a 200 amino acid sequence will result in a $200 x 22$ matrix as the node input and a $200 x 200$ adjacent matrix as edge input. The prediction labels were generated as in \cite{ref22}, where binary classification was performed on improved vs weakened binders, and regression prediction was performed on the free binding energy changes values. Figure \ref{fig:fig1} describes the modeling workflow for binary classification task.  

\begin{figure}
	\centering
	\includegraphics[width = 8.3cm]{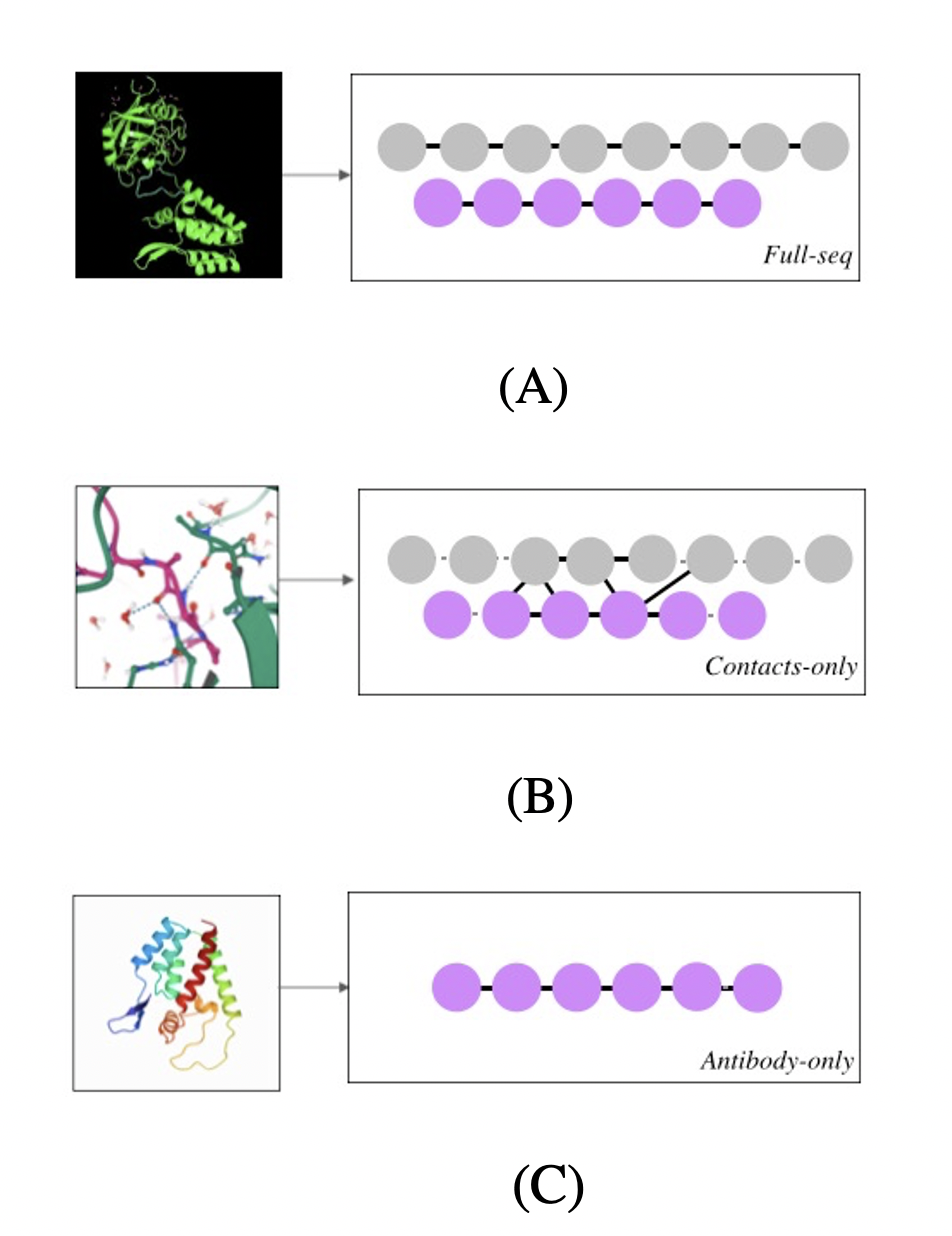}
	\caption{Graph representations of antibody-antigen interaction modeling.  (A) Full-seq graph (B) Contacts-only graph (C) Antibody-only graph}
	\label{fig:fig2}
\end{figure}

Following the above described workflow, we explore the graph representation of antibody-antigen complex with three different representation strategies.  1) Full-seq model: the full-seq model simply takes antibody and antigen sequences as two separated graph sequences(Figure \ref{fig:fig2}(A)). The intuition is to incorporate both interact contacts (ICs) and non-interacting surface(NIS) into modeling as the binding strength between antibody and antigen relies on the full conformation of the formed complex \cite{ref9,ref12} .  2) Contacts-only model: the contacts-only model produces a compact representation of the complex by utilizing residues on the interfacial surface (distance <5 Angstrom) only(Figure \ref{fig:fig2}(B)); Given limited high-quality training data, this approach is presumably more adequate since it models the most relevant information for antibody-antigen interactions. The identification of interface contacts was obtained using Prodigy-based prediction service\cite{ref9} based on complex’s 3D structure.  3) Antibody-only model: the antibody-only model aims to address the promiscuous binding capability of antibodies cross diverse antigens \cite{ref15,ref16,ref17,ref18}. We investigated on antibody-only modeling for binding affinity prediction, and evaluate if Hag-Net based network structure captures the enabling features for antibodies’ natural binding capability.

\subsubsection{Pairwise-study:\textit{ binding affinity pairwise rank prediction} }
Antibody maturation aims to optimize the binding affinity based on known antibody leads targeting specific antigens.  To this purpose, we construct the pairwise problem and study the binary relations between mutated variants of each therapeutic lead. Specifically, for each wildtype antibody, its associated mutations are grouped into unique ordered pairs. Each pair consists of two complexes with same target antigen but different mutated variants. If the first complex ranks higher than the second complex with respect to their binding affinities (i.e. the first complex possesses higher affinity), the pair is labeled as 1, otherwise as 0. Therefore, our goal is to obtain classifier f : 
$$
f(a,b) = \left\{ \begin{array}{ll}
1 & \textrm{if $\Delta\Delta G_a < \Delta\Delta G_b$}\\
0 & \textrm{else}\\
\end{array} \right.
$$

\begin{figure}
	\centering
	\includegraphics[width = 17cm]{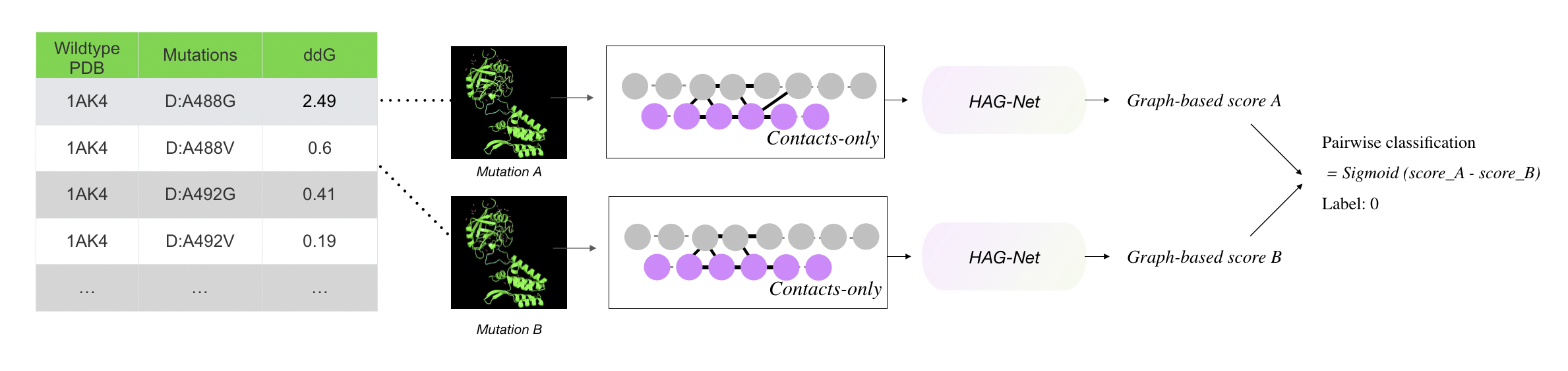}
	\caption{Pairwise classification model with contacts-only representation. In the presented table, each wildtype ( e.g. 1AK4) are listed with their associated mutated and binding affinity measurement. To construct an ordered pair, we take two mutations (mutation A and mutation B) and generate corresponding graph representations respectively. The outputs of Hag-Net are used as affinity scores for the comparison between mutations.  The difference between affinity scores then goes through sigmoid function to predict the binary relation label, specifically in this example, label is set to 0 since mutate A has lower binding affinity than mutate B.}
	\label{fig:fig3}
\end{figure}

Note that only mutations of the same therapeutic lead are ranked and grouped into ordered pairs. Our pairwise task is designed to predict the binary relations between antibody mutations, in order to evaluate if Hag-Net based model captures the enabling features that contribute to the affinity improvement upon mutations. By modeling the pairwise classification labels instead of binding free energy value, this approach also alleviates the possible errors and biases in the curated data due to different data sources and assays. Figure \ref{fig:fig3} illustrates the prediction workflow based on Hag-Net modeling for a single ordered pair of antibody complex 1BJ1.

\subsubsection{Benchmark study: \textit{LSTM vs Hag-Net} }
\begin{figure}
	\centering
	\includegraphics[width = 15.3cm]{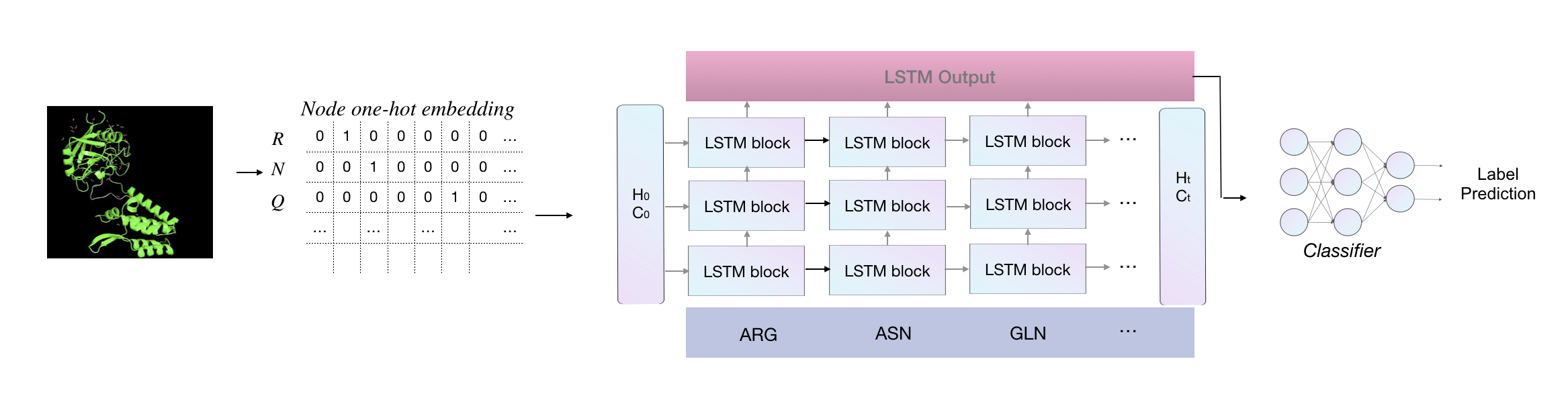}
	\caption{LSTM-based antibody-only classification modeling. Sequence information is converted to an input matrix by one-hot encoding and feed into a three-layer LSTM network.}
	\label{fig:fig4}
 \end{figure}

Finally, we compared the Hag-Net based network structure with the long short-term memory (LSTM) model, the latter was extensively used in natural language processing field and was previously examined in the research for protein sequence analysis [Mason et al]. Both models were trained using the same set of training data to assess their performance and scalability.  Figure \ref{fig:fig4} illustrates the LSTM approach for affinity prediction task.

\section{Results}

We investigated through different modeling approaches for in silico antibody maturation optimization .  Both five-folds and out-of-distribution cross-validation were performed to evaluate model capability as well as scalability. In the out-of-distribution experiments, five antigens were randomly selected from 32 antigens with their associated antibody-antigen complexes defined as the testing set (i.e. leave-five-out) , whereas the rest 28 antigens and associated complexes were used as the training set.  Performance of classification and regression were assessed by constructing ROC area under curve (AUC) and Pearson correlation coefficient respectively.

\subsection{Baseline study}
Table 1 shows the averaged AUC of binary classification on improved vs weakened binders using Hag-Net based models. Follow the protocol in \cite{ref22}, the testing set was split into subsets based on their $\Delta\Delta G$ absolute values (|$\Delta\Delta G$|>1, |$\Delta\Delta G$|>0.5 and |$\Delta\Delta G$|>0) and evaluated separately (Table 1). Benchmark studies performance from \cite{ref22}are also listed for comparison. As shown in Table1, the three Hag-Net modeling approaches are able to delineate improved and weakened binders well with respect to all subsets (Table1 (A)); In five-folds cross-validations, Hag-Net models yield satisfactory prediction accuracy in all subsets, and reveals high delineating capability on subsets with |$\Delta\Delta G$|>1 (with AUC > 0.9) ;

The capability of predicting ‘unseen’ complex structures is assessed by comparing the benchmark performance (Table1(A)) with out-of-distribution performance based on Hag-Net (Table1(B)). Specifically, for |$\Delta\Delta G$|>0, Hag-Net based approaches yield comparable delineation accuracy when comparing to the best performance reported (AUC = 0.73 from Discovery Studio); For samples with more significant changes in their binding affinities (|$\Delta\Delta G$|>0.5 and |$\Delta\Delta G$|>1), Hag-Net significantly outperforms conventional approaches. This demonstrates that Hag-Net based approaches can be used in predictions on completely unseen variants during antibody maturation. We hypothesized that Hag-Net captures the transferable features that contributes to the binding strength in antibody-antigen interactions.

We observed that the antibody-only modeling, when comparing to the other two approaches which incorporated antigen into modeling, shows solid delineating capability for weakened vs improved binders classifications, in both five-folds validations and novel variants predictions. This delineating capability observed in deep learning models is in line with antibodies naturally binding capability across antigens as previously reported  \cite{ref15,ref16,ref17,ref18}

 \begin{figure}
	\centering
	\includegraphics[width = 15.3cm]{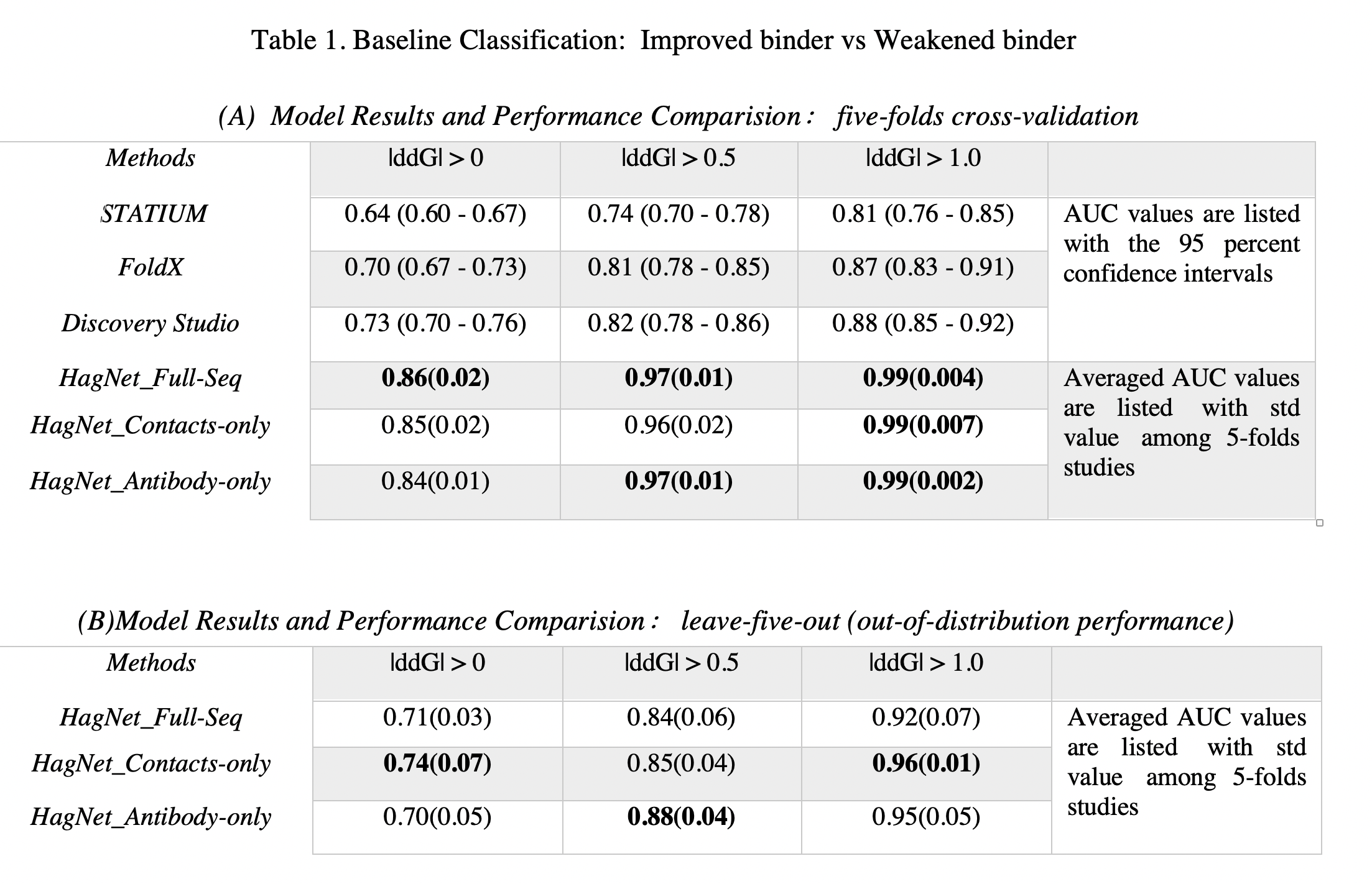}
	\label{fig:fig6}

 \end{figure}

Table 2 shows the regression performance using the metric of Pearson correlation coefficients. Full-seq and contacts-only approaches yield satisfactory regression performance in five-folds validations. $\Delta\Delta G$ value prediction on unseen variants are less significant yet still comparable with the best performance among conventional in silico approaches. We noticed that the antibody-only models possess less accuracy in both five-folds and out-of-distribution experiments, as it does not capture antibody-antigen specificity into modeling.

  \begin{figure}

	\centering
	\includegraphics[width = 15.3cm]{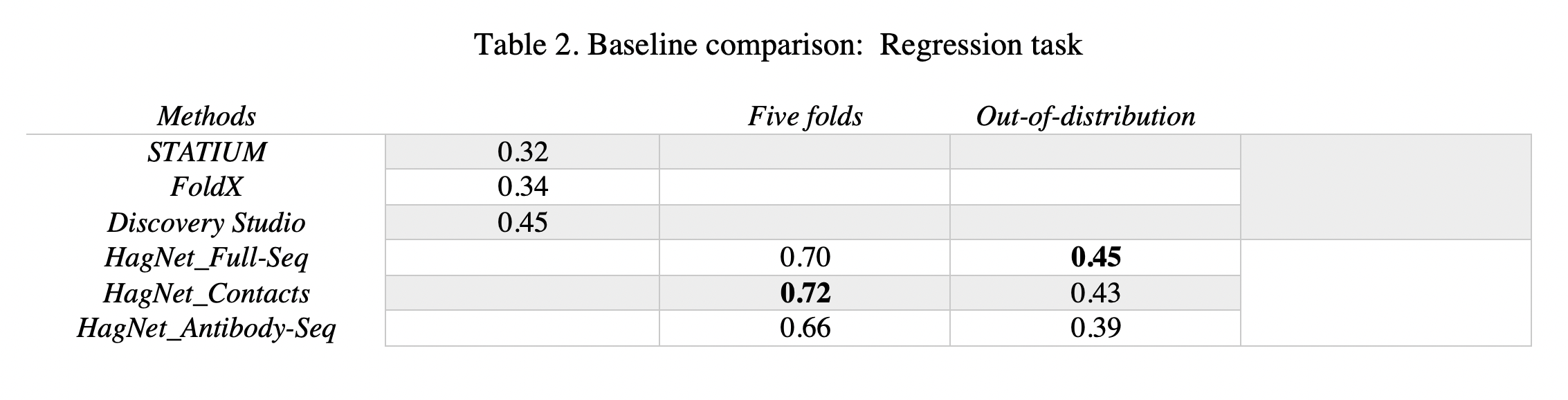}
	\label{fig:fig8}
 \end{figure}

\subsection{Pairwise Study}
Overall, 97711 pairs were generated from the original 1101 mutants of 32 complexes. The positive/negative ratio is 1 due to the intrinsic design for ordered pairs (i.e. for each positive pair, a negative pair was generated). We split the data into training and testing set (80/20).  Table 3 shows the pairwise task performance (AUC) with respect to full-seq, antibody-only and contacts-only representations. The pairwise problem aims to find favorable mutated variants of the same therapeutic lead with respect to their binding affinity, as opposed to baseline studies, in which different wildtypes and associated mutations are broadly defined as either improved or weakened binders. In essence, we anticipate the pairwise study to predict the proper ranks of the antigen-specific binding strength on mutated variants. 

Note that as the proposed pairwise study does not consider the affinities between different therapeutic leads (i.e. wildtypes), its results are not applicable to be compared with baseline studies. |$\Delta\Delta G$| subset does not apply in pairwise study either due to the intrinsic design of ordered pair generation. The averaged pairwise classification performance are presented in Table 3.

In Table 3, all three proposed graph representations achieved excellent results in five-folds studies, suggesting that Hag-Net based modeling is feasible to be used for the antibody maturation process. Specifically, full-seq modeling significantly outperforms the other two approaches in out-of-distribution experiments with AUC 0.70, indicating that both interacting contacts (ICs) and non-interacting surface (NIS) contribute to the affinity changes on therapeutic leads upon mutations.

 \begin{figure}
	\centering
	\includegraphics[width = 10.3cm]{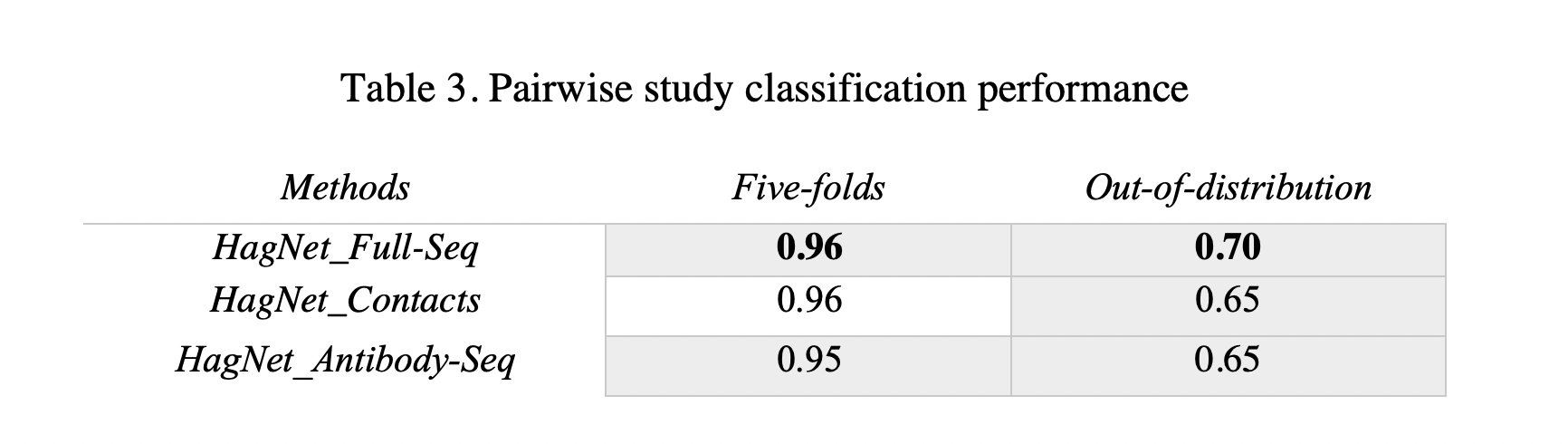}
	\label{fig:fig9}
 \end{figure}

\subsection{Deep Learning Benchmark Study}

To further evaluate the Hag-Net adequateness as the backbone structure for antibody-antigen interaction modeling, we incorporate the long short-term memory (LSTM) based modeling as an alternative approach and compare the two strategies with respect to training efficiency, prediction accuracy along and generalization capability. Antibody-only representation was employed for model simplicity. Figure 5 depicts the convergence curves of LSTM based and Hag-Net based training on one of five folds during cross-validations in pairwise study.

\begin{figure}
 	\centering
	\includegraphics[width = 15.3cm]{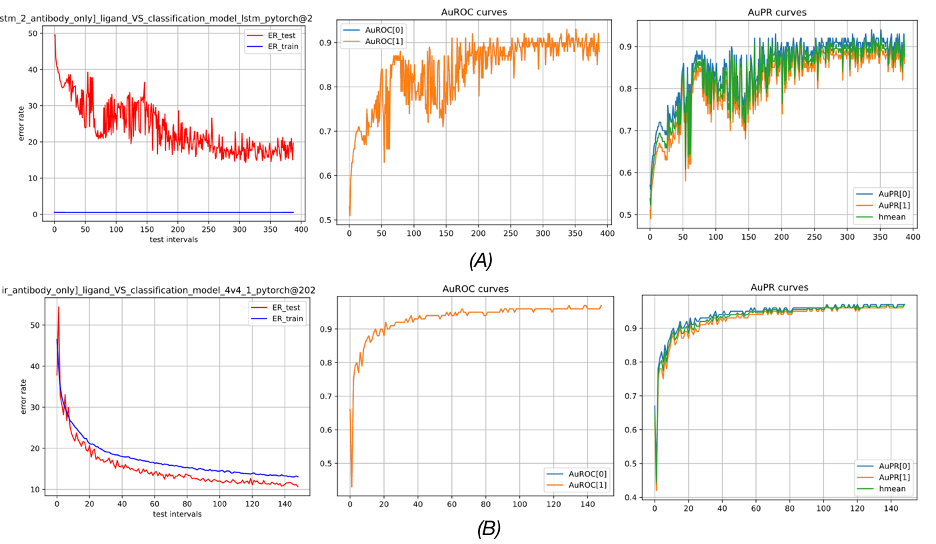}
	\caption{ Convergence curves of error rate(left), AUC (middle) and AUPR(right) for class 1 and class 0 on pairwise dataset (A) LSTM-based model (B) Hag-Net based model }
	\label{fig:fig5}
 \end{figure}

 In Figure \ref{fig:fig5}, we observed that Hag-Net converges faster than LSTM during training with higher prediction accuracy (performance showed LSTM at epoch 400 and Hag-Net at epoch 150), and results in more stable and consistent training performance. Table 3 demonstrates the prediction accuracy of the two approaches , where Hag-Net model yields better delineation in five-folds analysis and lesser performance in out-of-distribution tasks. 

 \begin{figure}
	\centering
	\includegraphics[width = 10.3cm]{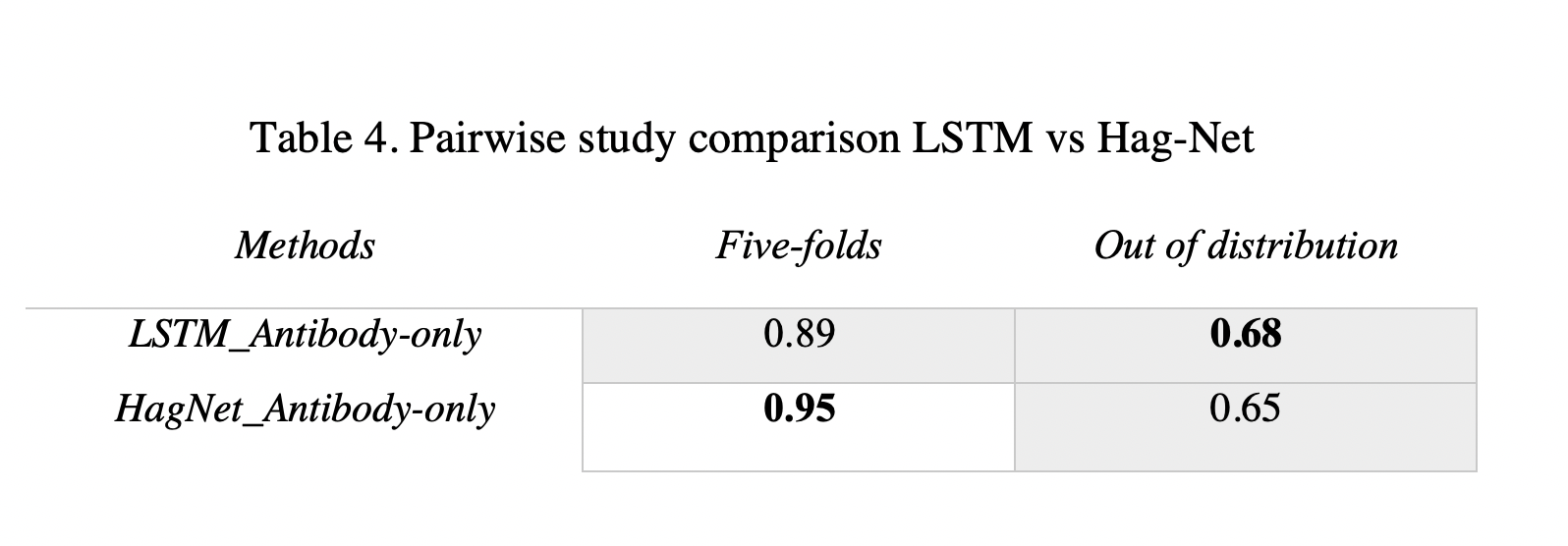}
	\label{fig:fig10}
 \end{figure}

\section{Discussions}
In the last several decades, the number of known protein sequences has grown exponentially along with the development of sequencing technologies, yet the number of corresponding crystal structures is limited due to their high dependency on experiments. To examine the potential of sequence data- guided in silico antibody maturation, we proposed affinity prediction models that utilize deep learning techniques based on the complex sequence information only. The intuition of such modeling strategy is to capture the enabling features encrypted at amino acids level that contributes to the interactions and the resulting binding strength between antibody and antigen.  By modeling at the level of amino acids level instead of atoms, the proposed approach significantly reduces the time cost for both training and prediction process, while possesses better tolerance on the size of annotated training datasets due to the model’s simplicity. This structure-free optimization approach also allows it applicability on broader scope of analysis given the inevitable disparity between available sequence data and crystal structures in antibody and protein research field.

Since the discovery of antibodies, researcher have been searching for the interpretation of relationship between limited number of antibodies and their binding ability to the almost infinite antigens space \cite{ref15}. Despite antibodies’ exquisite specificity to target antigens, studies have reported and examined antibodies promiscuous binding capabilities \cite{ref15}.   \cite{ref18} postulated that the presence of antibodies’ poly-reactivity may be caused by the flexible antigen-binding ‘pocket’, which can change conformation to accommodate different antigens. Such cross-reactivity is usually studied as a challenge for high specificity antibody development \cite{ref17}, while on the other hand it also allows the invention on broad-specificity antibody thru engineering approaches \cite{ref16}. In our antigen-independent experiments, antibody-only models have exhibited solid prediction capability for the delineations between improved vs weakened binders,  and can be further applied on unseen and novel samples outside the training distribution. This might suggest that such the ‘natural binding’ capability possessed by antibodies can be modeled by graph-based network structure with proper model designs, and ought to be further investigated for antibody maturation development with stronger data diversity and larger sample size.

With the propagation between graph-convolutional layers and pyramid features intrinsically designed in the graph-based structure, the well-trained Hag-Net models are expected to represent the antibody-antigen complex in a much more comprehensive fashion comparing to traditional machine learning approaches. This is demonstrated by the higher accuracy (AUC 0.95 vs AUC 0.89) and significantly faster and smoother training achieved in Hag-Net when comparing to LSTM models (Figure \ref{fig:fig5}). We notice, however, that the Hag-Net model inevitably suffers from larger number of parameters due to its structure and is therefore inclined to the overfitting problem given limited training data, resulting in the lesser generalization performance in out-of-distribution tasks.

Through our analysis, the prediction performance and model scalability are inevitably limited by the size of publicly available dataset with annotated binding affinity. The proposed method and trained models ought to be validated on larger dataset, and its feasibility in assisting the antibody maturation need to be further evaluated along with industry practice  on unseen wildtype samples. We are also aware that the proposed modeling for antibody-antigen interactions does not guarantee its applicability in other protein-protein interactions (PPI), as antibodies possess unique properties and does not fall into the general protein interaction space.

\section{Conclusion}
We proposed Hag-Net based deep learning modeling approaches for antibody maturation based on sequence data. The contributions on interacting contacts (ICs), non-interacting surface (NIS) in terms of affinity prediction are studied through different modeling designs, along with the antibodies’ naturally binding capability. The proposed approaches achieved excellent performance in both five-fold cross-validations and out-of-distribution analysis, and outperformed conventional in-silico analysis on AB-BIND dataset. Our pairwise study further investigated Hag-Net models’ feasibility in studying the antigen-specific binding strength under the therapeutic leads optimization scenario. In the end, we performed benchmark comparison by using the LSTM-based backbone structure for modeling, assessing training efficiency as well as model performance between language modeling and graph-based modeling in the application of antibody-antigen analysis. We concluded that the proposed methodology, though limited by sample size, possesses high efficiency and robust delineation capability for antibody affinity prediction and therefore holds great potential to be further validated and utilized in the antibody development pipeline \cite{ref1}.

\section{Acknowledgement}
The 3D structures of antibody complex 1BJ1 in Figure 1 - 4 are obtained from \cite{ref19} and generated using molecular visualization system Pymol  \cite{ref21}

\end{document}